\documentclass[%
 notitlepage,
 reprint,
 superscriptaddress,
 amsmath,amssymb,
 aps,
]{revtex4-2}

\usepackage{hyperref}
\hypersetup{colorlinks=true,linkcolor=blue,citecolor=blue,urlcolor=blue}
\usepackage{lipsum}
\usepackage{graphicx}
\usepackage{dcolumn}
\usepackage{bm}

\usepackage{epstopdf}
\usepackage{xcolor}
\usepackage{siunitx}
\begin{document}

\preprint{APS/123-QED}

\title{Nonlinear dynamics and self-healing properties of elliptical Airy beams in Kerr media}

\author{Qinjun Jian}
 \altaffiliation{These authors contributed equally to this work.}
\author{Jing Hu}
 \altaffiliation{These authors contributed equally to this work.}
 \affiliation{
Key Laboratory for Physical Electronics and Devices of the Ministry of Education \& Shaanxi Key Lab of Photonic Technique for information, School of Electronics Science \& Engineering, Faculty of Electronic and Information Engineering, Xi’an Jiaotong University, Xi’an, 710049, China}%
\author{Lihe Yan}%
 \email{liheyan@mail.xjtu.edu.cn}
 \affiliation{
Key Laboratory for Physical Electronics and Devices of the Ministry of Education \& Shaanxi Key Lab of Photonic Technique for information, School of Electronics Science \& Engineering, Faculty of Electronic and Information Engineering, Xi’an Jiaotong University, Xi’an, 710049, China}%
 \affiliation{%
Engineering Research Center of Integrated Circuit Packaging and Testing, Ministry of Education, Tianshui Normal University, Tianshui, 741000, China}
\author{Jinhai Si}
\author{Xun Hou}
\affiliation{
Key Laboratory for Physical Electronics and Devices of the Ministry of Education \& Shaanxi Key Lab of Photonic Technique for information, School of Electronics Science \& Engineering, Faculty of Electronic and Information Engineering, Xi’an Jiaotong University, Xi’an, 710049, China}%

\begin{abstract}
By numerically solving the nonlinear Schrödinger equation, we theoretically study the nonlinear propagation dynamics and self-healing properties of elliptical Airy beams (EABs) propagating in water under Kerr nonlinearity. Compared to linear propagation, EABs exhibit extended propagation distances and enhanced stability in nonlinear media. Furthermore, particular emphasis is placed on the impact of Kerr nonlinearity strength on the propagation and self-healing properties of EABs. By varying the input power, it is found that EABs within a moderate power range can propagate longer distances while maintaining higher intensity and exhibit improved robustness after being blocked, indicating better self-healing performance. Based on this analysis, we propose an optimal input power for EABs through a quantitative analysis of the impact of Kerr nonlinearity,  enabling them to achieve the greatest propagation distance and maintain the highest stability. Our work provides a comprehensive theoretical understanding of the nonlinear propagation dynamics and self-healing properties of EABs, with their superior characteristics potentially applicable to long-distance laser transmission.
\end{abstract}

\maketitle

\section{Introduction}

Nondiffracting beams, characterized by the self-healing properties and capacity to maintain intensity profiles over long propagation distances, have found extensive applications in various fields \cite{yun2024high, zhang2019robust, zhang2021generation}. In this context, Airy beams have been widely studied \cite{jia_isotropic_2014, vettenburg_light-sheet_2014, ellenbogen_nonlinear_2009, polynkin_filamentation_2009} due to their unique non-diffractive, self-bending \cite{siviloglou_observation_2007,siviloglou_accelerating_2007}, and self-healing \cite{broky_self-healing_2008} properties since their first experimental demonstration in 2007 \cite{siviloglou_observation_2007,siviloglou_accelerating_2007}. Subsequently, the circular Airy beam (CAB), formed by radially symmetric Airy beams, has also attracted great interest since its proposal and realization \cite{efremidis_abruptly_2010, papazoglou_observation_2011} for its particular propagation dynamics, resulting in a variety of applications \cite{lu_abruptly_2019, manousidaki_abruptly_2016, lu_dynamical_2021, panagiotopoulos_sharply_2013, hu_metasurface-based_2024, cai_dynamic_2020, chen_optimizing_2024}. For instance, its abruptly autofocusing property has facilitated its application in optical manipulation \cite{lu_abruptly_2019, manousidaki_abruptly_2016, lu_dynamical_2021}, and its extended propagation distance and self-healing property have rendered it particularly suitable for optical bullets \cite{panagiotopoulos_sharply_2013} and long-distance optical communication \cite{hu_metasurface-based_2024}. Similarly, the elliptical Airy beam (EAB) \cite{zha_elliptical_2018, xie_propagation_2018, cao_characteristics_2020, chen_elliptical_2024, he_realization_2024}, characterized by its asymmetric field distribution, not only retains the advantageous traits of CABs, including abruptly autofocusing \cite{zha_elliptical_2018} and self-healing capabilities, but also exhibits a more extended propagation distance \cite{xie_propagation_2018}. These characteristics facilitate enhanced control over transmission, endowing EABs with distinctive advantages in various potential applications, particularly in long-distance and stable laser transmission.

When high-power lasers propagate in nonlinear media, the propagation properties are significantly affected by Kerr nonlinearity \cite{panagiotopoulos_sharply_2013, chen_effect_2010, polynkin_filamentation_2009,panagiotopoulos_sharply_2013, chen2023controllable, kaminer_self-accelerating_2011, allayarov_dynamics_2014, bouchet2018solitonic}. More specifically for Airy beams, the nonlinear effects result in field deformation \cite{chen_effect_2010, chen2023controllable} and soliton emission \cite{kaminer_self-accelerating_2011} during the propagation, and also significantly impact their self-healing properties \cite{abdollahpour_spatiotemporal_2010}. Another example occurs for CABs in a nonlinear regime, where the nonlinear effects reshape the optical wavepacket, resulting in the formation of a long and uniform filament \cite{panagiotopoulos_sharply_2013}. In conclusion, it was found that Kerr nonlinearity exerts a distinct influence on the propagation characteristics and self-healing properties of Airy beams and CABs, depending on the different field distributions between the beams. Building on this, it is essential to investigate the specific impact of Kerr nonlinearity on EABs, considering their unique propagation properties and asymmetric field distribution. A key question is whether these beams can maintain their self-healing properties over extended propagation distances in nonlinear media, such as water. However, research on the nonlinear dynamics of EABs remains limited, despite their potential for long-distance structured light propagation. Consequently, a deeper understanding of the impact of Kerr nonlinearity on the propagation characteristics and self-healing properties of EABs is crucial for optimizing their performance in nonlinear media.

In this paper, we study the nonlinear propagation characteristics and self-healing properties of EABs in water by numerically solving the nonlinear Schrödinger equation. We first investigate the propagation dynamics of EABs and demonstrate their extended propagation distances in nonlinear media compared to those in linear media. Furthermore, by varying the input power, we explore the influence of Kerr nonlinearity on the propagation characteristics of EABs. It is found that the beams achieve optimal propagation distance and peak intensity under moderate nonlinearity conditions. Next, we explore the self-healing properties of EABs, showing that the beams can be reconstructed and propagate stably over longer distances even after being partially obstructed. The robustness of EABs under weak and moderate nonlinearity in obstacle environments is further confirmed, highlighting their potential for reliable performance in complex environments. Finally, an optimal input power that maximizes both the propagation distance and self-healing properties of EABs is proposed by quantitatively analyzing the effect of Kerr nonlinearity. The extended propagation distance and stability of high-intensity EABs in water are expected to find applications in the context of underwater long-distance transmission.

The paper is organized as follows. The theoretical model adopted in this study is presented in Sec.\,\ref{section 2}. In Sec.\,\ref{section 3}, we discuss the effects of Kerr nonlinearity on the nonlinear propagation dynamics of EABs in pure water. Section\,\ref{section 4} investigates the self-healing properties of EABs in obstacle environments. Then, we propose an optimal input power for EABs propagation in water by analyzing the effect of Kerr nonlinearity in Sec.\,\ref{section 5}. Lastly, Sec.\,\ref{section 6} summarizes the results of our work.

\section{Theoretical model}
\label{section 2}
The principal features of high-intensity EAB evolution in kerr madia, including diffraction and nonlinear Kerr effects, can be described in the paraxial approximation by the nonlinear Schrödinger equation \cite{chen_effect_2010}:
\begin{equation}
\frac{\partial A}{\partial z} = \frac{i}{2k}\nabla^2_{\bot} + ik\frac{n_2}{n_1} \left| A \right| ^2 A,
\end{equation}
where the terms on the right side of Eq. (1) describe the transverse diffraction and the Kerr nonlinearity in turn. 
 $\nabla^2_{\bot} = \partial^2/\partial x^2 + \partial^2/\partial y^2$ is the two-dimensional transverse Laplace operator, and $k = 2\pi n_0/\lambda$ is the wavenumber of the optical wave, $n_0$ and $n_2$ is the linear and nonlinear refractive index, respectively.

Referring to the expression of initial field distribution of the CAB \cite{papazoglou_observation_2011}, the electric field distribution of an EAB in the initial plane $z = 0$ is described by
\begin{equation}
\begin{aligned}
A(x,y,0) = &A_0 \cdot Ai(\frac{r_0-\sqrt{(x/t)^2+y^2}}{\omega_0})
\\&\times\exp{\left [(\alpha + ic_1) \frac{r_0-\sqrt{(x/t)^2+y^2}}{\omega_0} \right ]},
\end{aligned}
\end{equation}
where $\omega_0$ is a scaling factor, $r_0$ is the radius of the primary ring, $\alpha$ is an exponential decay factor, $c_1$ is a linear chirp factor that increases the focal length of the EAB \cite{he_key_2023}, and $t$ is the elliptical parameter, indicating the ratio of the short axis (x-axis) of an ellipse to the long axis (y-axis), and for $t = 1$ , EAB degenerates to CAB \cite{papazoglou_observation_2011}. The coefficient $A_0$ can be derived by using the relationship $P = \int^{\infty}_{-\infty} \int^{\infty}_{-\infty} \left | A \right | ^2 dx dy$ , which is expressed as
\begin{equation}
A_0 = \sqrt{\frac{P}{t \omega_0^2 \sqrt{\pi/2\alpha}} \cdot \exp{(\frac{2\alpha^3}{3})} \cdot (\frac{r_0}{\omega_0} + \frac{1-4\alpha^2}{\alpha} )} \cdot \sqrt{\frac{1}{t}},
\end{equation}
with $P$ being the initial beam power. When $t = 1$ , Eq. (3) can be reduced to that of CABs, which is consistent with the result presented in Ref. \cite{efremidis_abruptly_2010}.

In this paper, we use the split-step Fourier method \cite{agrawal_nonlinear_2013} to solve the nonlinear Schrödinger equation numerically, and the main code is provided in the Supplemental Material. Unless otherwise specified, the numerical calculation parameters for EABs are taken as $\lambda = \SI{532}{nm}$, $\omega_0 = \SI{200}{\mu m}$, $r_0 = \SI{2}{mm}$, $t = 0.8$, $\alpha = 0.2$ and $c_1 = -0.6$, propagating in a general nonlinear medium (water) with $n_0 = 1.33$ and $n_2 = 4.1 \times 10^{-16} ~\textup{cm}^2/\textup{W}$ \cite{boyd_nonlinear_2008}. In addition, the critical power for self-focusing of the Gaussian beam $P_{\textup{cr}} = 3.77\lambda^2 / 8 \pi n_0 n_2 \approx \SI{0.78}{MW}$ \cite{panagiotopoulos_sharply_2013} is utilized as the power normalization parameter in this paper.

\section{Nonlinear propagation dynamics of EABs in pure water}
\label{section 3}
\subsection{Propagation properties of EABs}
\begin{figure*}[htbp]
\centering\includegraphics[width=1.65\columnwidth]{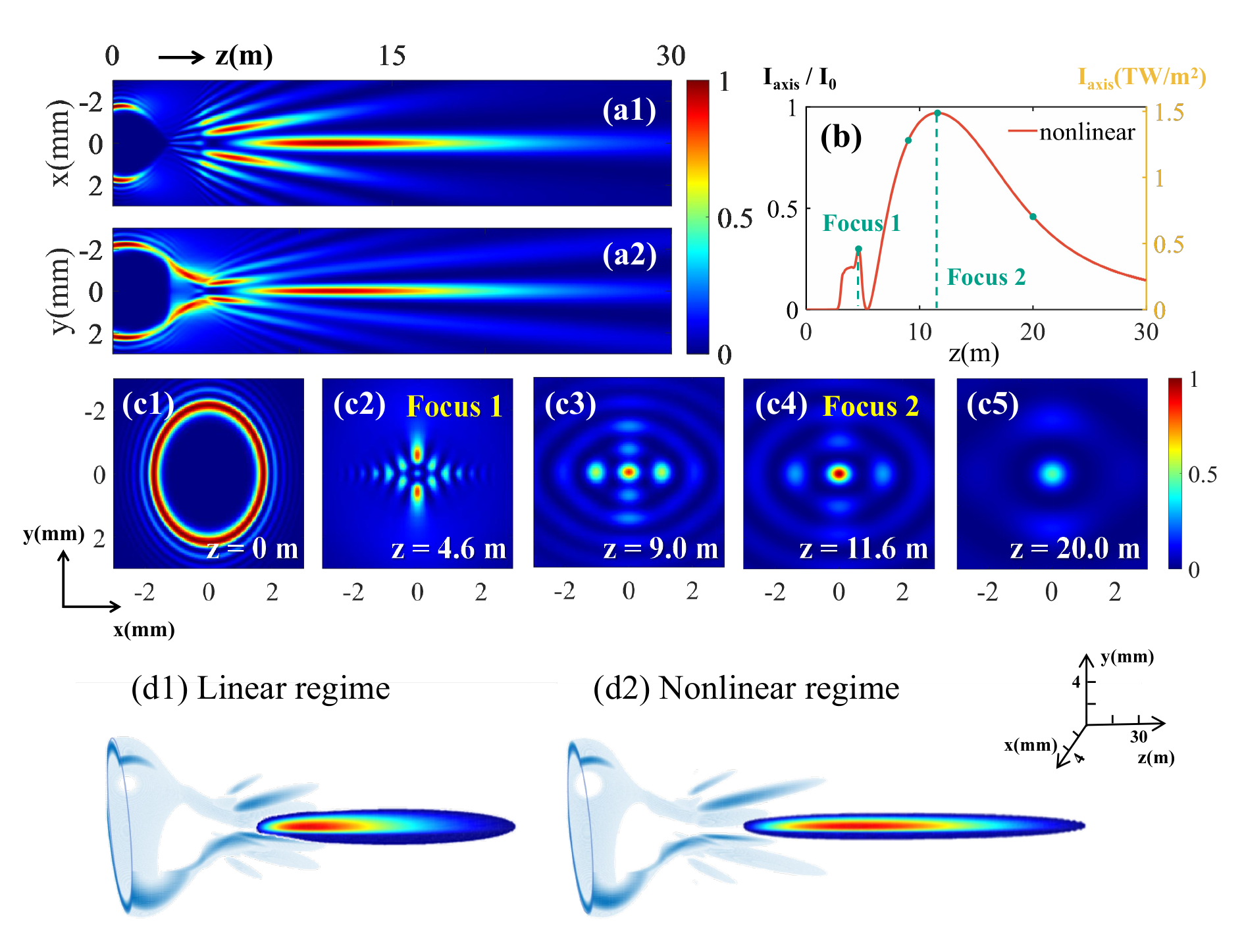}
\caption{The intensity distributions of an EAB in pure water.  (a) Longitudinal intensity distribution in the (a1) x-z plane and (a2) y-z plane. (b) The normalized intensity $I_{\textup{axis}}/I_0$ (left y-axis) and intensity $I_{\textup{axis}}$ (right y-axis) along the z-axis versus the propagation distance z, and the green scatters correspond in turn to the different propagation distances in (c1-c5). (c) Transverse intensity profiles at the propagation distance (c1) $z = \SI{0}{m}$, (c2) $z = \SI{4.6}{m}$ (Focus 1), (c3) $z = \SI{9}{m}$, (c4) $z = \SI{11.6}{m}$ (Focus 2), and (c5) $z = \SI{20}{m}$. (d) Graphic representation of EABs in the (d1) linear and (d2) nonlinear regime.}
\label{fig1} 
\end{figure*}

In this section, based on the above theoretical model, the field distribution of EABs is first simulated to show the nonlinear propagation characteristics. The normalized intensity distributions $I/I_0$ of the EAB with an input power $P_{\textup{in}} = 10 P_{\textup{cr}}$ in pure water are analyzed as shown in Fig.\,\ref{fig1}, where $I_0$ is the peak intensity of the initial field. The nonlinear longitudinal intensity distribution, on-axis intensity, and transverse intensity profile at different propagation positions are shown in Fig.\,\ref{fig1}(a-c), respectively. Moreover, Fig.\,\ref{fig1}(d) illustrates a comparison between linear and nonlinear propagation, with only the diffraction effect considered in the linear case \cite{zha_elliptical_2018}.

From Fig.\,\ref{fig1}(a) we can find that the EAB exhibits an abruptly autofocusing property and subsequently forms a uniform beam along the beam axis in the nonlinear case, which is similar to the characteristics observed in the linear case \cite{zha_elliptical_2018, xie_propagation_2018}. Two focuses are observed along the beam axis (as illustrated in Fig.\,\ref{fig1}(b)), resulting from the difference in length between the long axis (y-axis) and the short axis (x-axis) of the EAB. In nonlinear propagation, the energy of the short axis converges toward the center firstly, forming a lower-intensity focal point at Focus 1, as shown in Fig.\,\ref{fig1}(c2). Then, the energy of the long axis flows toward the center and the direction of short axis, resulting in the formation of a high-intensity elliptical focal point at Focus 2, as shown in Fig.\,\ref{fig1}(c4). At last, energy diverges outward slightly, forming a long and uniform beam along the beam axis. To visualize this dynamic process more clearly, we include Supplementary Video 1, which shows the evolution of the optical field during the propagation of EABs in nonlinear media. Furthermore, while the propagation of EABs in nonlinear media displays similarities with those in linear media, it notably exhibits a more uniform on-axis beam profile and a smaller beam width over longer propagation distances compared to the linear case, as depicted in Fig.\,\ref{fig1}(d). This phenomenon can be explained in terms of the Kerr nonlinearity, which reduces the influence of the diffraction effect and enables the on-axis beam to propagate over longer distances without significant diffraction.

\subsection{Effect of Kerr nonlinearity}
\begin{figure*}[tp]
\centering\includegraphics[width=1.7\columnwidth]{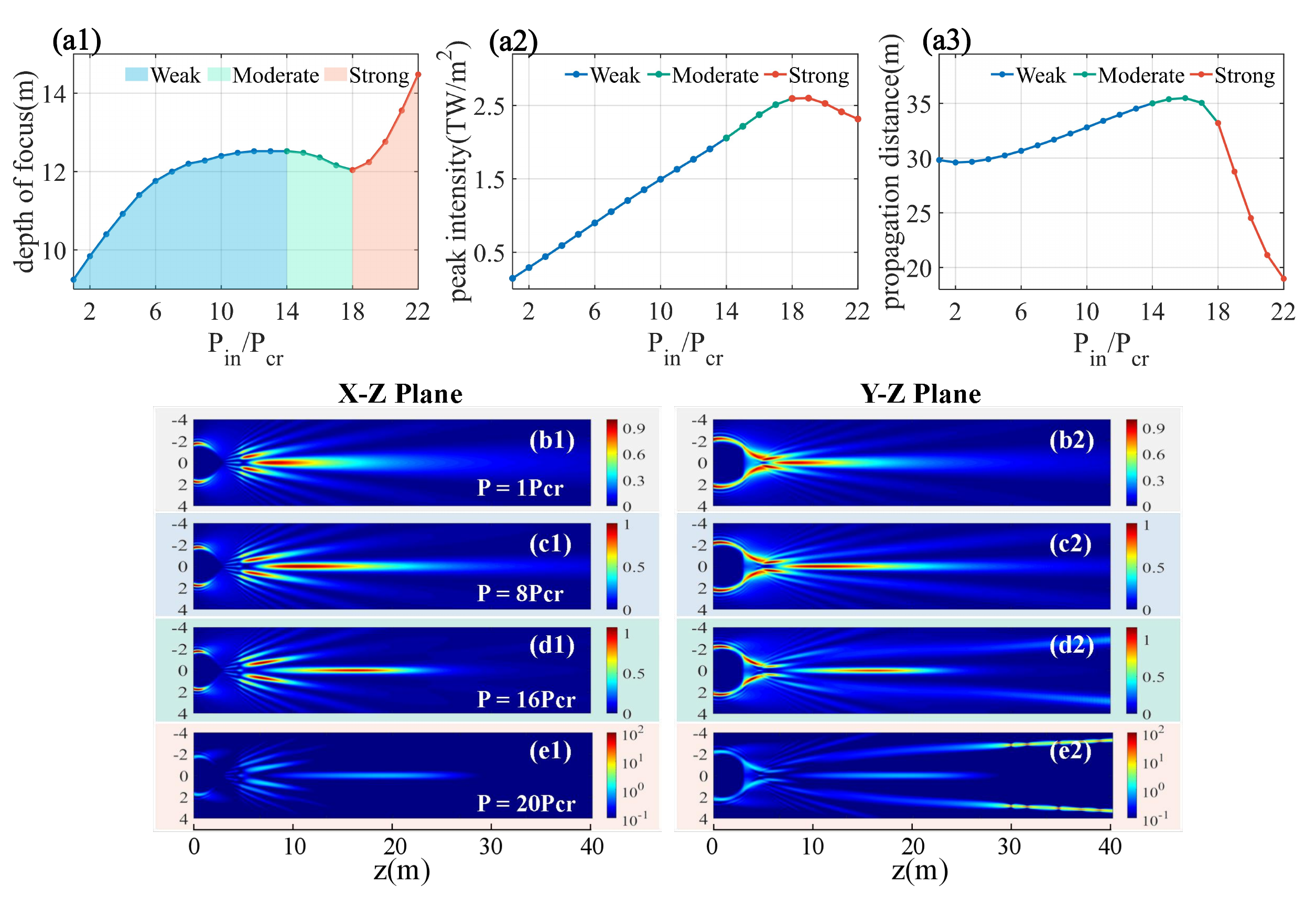}
\caption{Propagation dynamic of an EAB with different input powers. (a1) depth of focus (FWHM), (a2) Peak intensity of Focus 2, (a3) Propagation distance. (b-d) Longitudinal intensity distribution of an EAB with different input powers. (b) $P_{\textup{in}} = 1P_{\textup{cr}}$, (c) $P_{\textup{in}} = 8P_{\textup{cr}}$, (d) $P_{\textup{in}} = 16P_{\textup{cr}}$, (e) $P_{\textup{in}} = 20P_{\textup{cr}}$. The left column is the longitudinal intensity distribution in the x-z plane, while the right column presents the same data in the y-z plane.}
\label{fig2}
\end{figure*}

To further explore the effect of Kerr nonlinearity on the propagation properties of EABs in kerr media, we analyze the variations in the depth of focus (DOF), peak intensity $I_\textup{max}$ and propagation distance as functions of the input power, as illustrated in Fig.\,\ref{fig2}(a). Additionally, we present a series of specific longitudinal intensity distributions of underwater propagation, as shown in Fig.\,\ref{fig2}(b). Here, the depth of focus is defined as the full width at half maximum (FWHM) of the intensity $I_{\textup{axis}}$ along the z-axis \cite{chen_optimizing_2024}, and the propagation distance refers to the position where the power decays to $1/e^2$ of the input power within a $\SI{5}{mm} \times \SI{5}{mm}$ area centered on the beam axis.

Based on the distinct propagation properties of EABs, namely the different trends of the DOF (as shown in Fig.\,\ref{fig2}(a1)), the influence of Kerr nonlinearity can be classified into three regimes \cite{kaminer_self-accelerating_2011, efremidis_airy_2019}: weak nonlinearity for $P_{\textup{in}} < 14P_{\textup{cr}}$ (Fig.\,\ref{fig2}(b-c)), moderate nonlinearity for $14P_{\textup{cr}} \leq P_{\textup{in}} < 18P_{\textup{cr}}$ (Fig.\,\ref{fig2}(d)), and strong nonlinearity for $P_{\textup{in}} \geq 18P_{\textup{cr}}$ (Fig.\,\ref{fig2}(e)). In the case of weak nonlinearity (Fig.\,\ref{fig2}(b-c) and the blue line in Fig.\,\ref{fig2}(a1)), the DOF increases with increasing input power and reaches its maximum value at $P_{\textup{in}} = 14P_{\textup{cr}}$, as the self-focusing effect is enhanced by the stronger on-axis beam intensity. In the case of moderate nonlinearity (Fig.\,\ref{fig2}(d) and the green line in Fig.\,\ref{fig2}(a1)), the DOF decreases due to the competition between the self-focusing and diffraction effects. More specifically, the beam width is reduced by the self-focusing effect that causes the energy to converge toward the center. Then, the impact of the self-focusing effect decreases in the sense that the diffraction effect is enhanced by the smaller beam width, leading to a slight decrease in the DOF under moderate nonlinearity compared to the weak nonlinearity. In the case of strong nonlinearity (Fig.\,\ref{fig2}(e) and the red line in Fig.\,\ref{fig2}(a1)), the emission of solitons from EABs in the y-z plane and the interaction between the on-axis beam and these solitons lead to an increase in the DOF. However, the on-axis beam under strong nonlinearity is unstable due to the fact that the majority of the energy is concentrated in the soliton \cite{kaminer_self-accelerating_2011}. In conclusion, for weak or moderate nonlinearity, EABs exhibit stable propagation, whereas for strong nonlinearity, the propagation stability is reduced by the emission of solitons, which is consistent with the effect of Kerr nonlinearity on Airy beams reported in previous studies \cite{kaminer_self-accelerating_2011, efremidis_airy_2019}.

Besides, both the peak intensity $I_\textup{max}$ (as shown in Fig.\,\ref{fig2}(a2)) and the propagation distance (as shown in Fig.\,\ref{fig2}(a3)) show an initial increase followed by a decline with increasing power. In the case of weak nonlinearity, both the peak intensity and propagation distance increase with increasing Kerr nonlinearity, subsequently reaching a maximum in the moderate nonlinearity regime. However, when the input power reaches the state of strong nonlinearity, the majority of the energy is transferred into solitons, leaving only a minor portion in the main lobe along the beam axis \cite{allayarov_dynamics_2014}, resulting in a decrease in the peak intensity and propagation distance. In summary, it can be observed that the EAB under moderate nonlinearity propagates over a longer distance and with a higher intensity when considering the peak intensity, and propagation distance trends.

\section{Nonlinear propagation dynamics of EABs with obstacles}
\label{section 4}
In the previous section, it was demonstrated that the peak intensity and propagation distance of the EABs are increased by an appropriate input power in pure water. However, during practical underwater laser transmission, factors such as turbid water and the presence of underwater organisms can interfere with the propagation of laser beams, thereby affecting beam quality and causing field deformation. Herein, we investigate the self-reconstruction capabilities of EABs with one, two, and a large number of random obstacles in the propagating path, with a particular emphasis on the influence of Kerr nonlinearity on the propagation.
\subsection{Self-reconstruction of EABs}

\begin{figure*}[tp]
\centering\includegraphics[width=1.9\columnwidth]{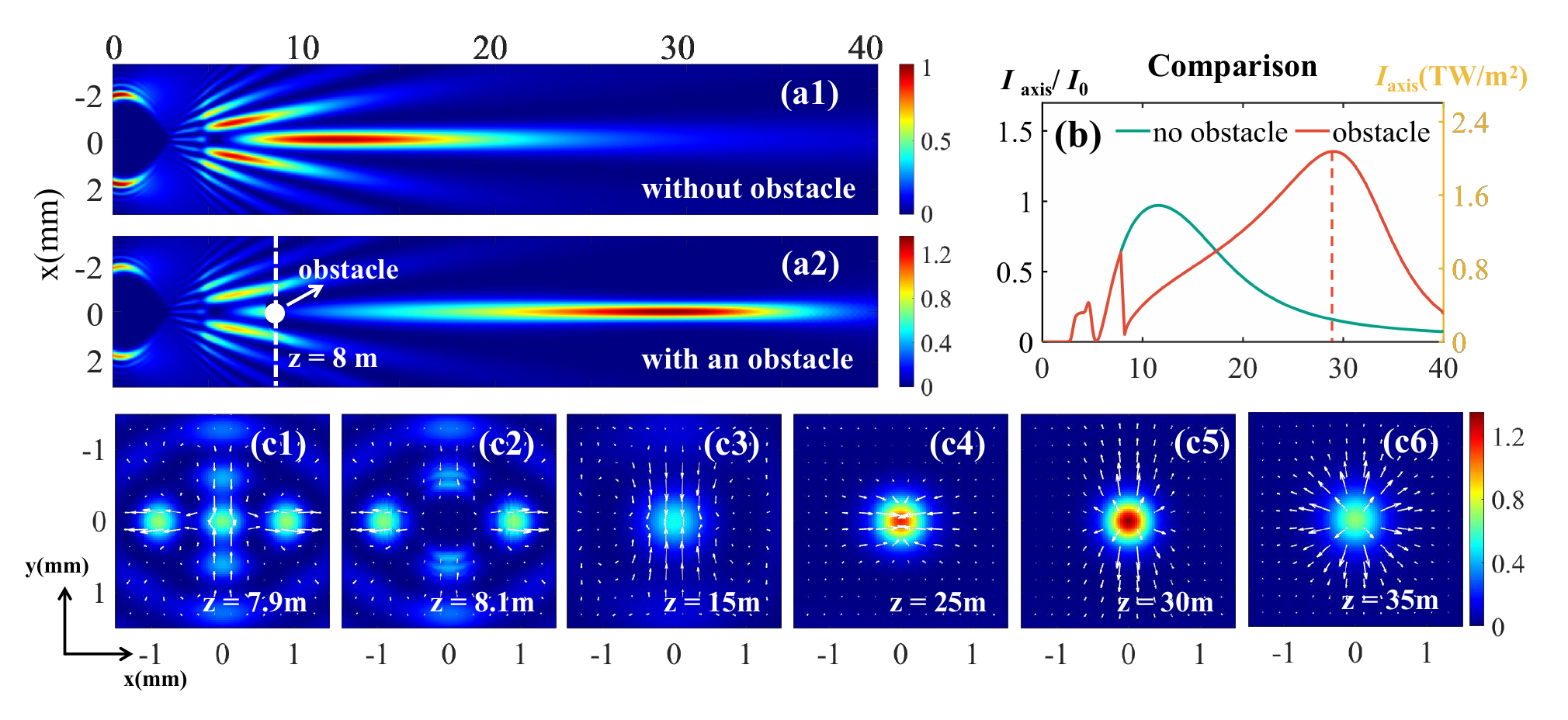}
\caption{Self-reconstruction of an EAB in water. (a) Longitudinal intensity distribution in the x-z plane (a1) without obstacle and (a2) with an obstacle. (b) the on axis intensity comparison between without obstacle (green line) and with an obstacle (red line). Transverse intensity profiles and  transverse power flow $\vec{S}_{\bot}$ (white arrows) at different propagation distance (c1) $z = \SI{7.9}{m}$, (c2) $z = \SI{8.1}{m}$, (c3) $z = \SI{15}{m}$, (c4) $z = \SI{25}{m}$, (c5) $z = \SI{30}{m}$ and (c6) $z = \SI{35}{m}$. }
\label{fig3}
\end{figure*}

In order to demonstrate the self-healing properties of the EAB with obstacle, we calculated power flow of EABs with obstacle by transverse Poynting vector $\vec{S}_{\bot}$ \cite{broky_self-healing_2008}:
\begin{equation}
\vec{S}_{\bot} = \frac{i}{4\eta k} \left [ A \nabla_{\bot} A^* - A^* \nabla_{\bot} A \right ]
\end{equation}
where $\eta = \sqrt{\mu_0/n_0^2 \epsilon_0}$ is the impedance in water, $\mu_0$ and $\epsilon_0$ are permeability and permittivity in a vacuum. To demonstrate the self-reconstruction of the EAB, a circular obstacle with a diameter of \SI{0.7}{mm} was placed at a propagation distance of $z = \SI{8}{m}$. As shown in Fig.\,\ref{fig3}, the longitudinal and transverse intensity distributions of the EAB with input power $P_{\textup{in}} = 10P_{\textup{cr}}$ are plotted. It can be observed that the beam gradually reconstructs and returns to its original transverse intensity profile before being blocked as the propagation distance increases when the main lobe of the EAB is blocked. In addition, we can find that the beam exhibits uniform propagation over a longer distance (Fig.\,\ref{fig3}(a1) and (a2)) and an increase in the on-axis intensity (Fig.\,\ref{fig3}(b)) when the main lobe is blocked by the comparison. From the perspective of power flow, as shown in Fig.\,\ref{fig3}(c) and Supplementary Video 2, power flows from the side lobes of the EAB toward the center to facilitate self-healing after the main lobe is blocked. It is the increased power flowing from the side lobes toward the center that causes the main lobe of EABs to maintain a higher on-axis intensity and propagate over a longer distance than in the unblocked case. 

\subsection{Two obstacles on the beam axis}
\begin{figure}[tp]
\centering\includegraphics[width=0.95\columnwidth]{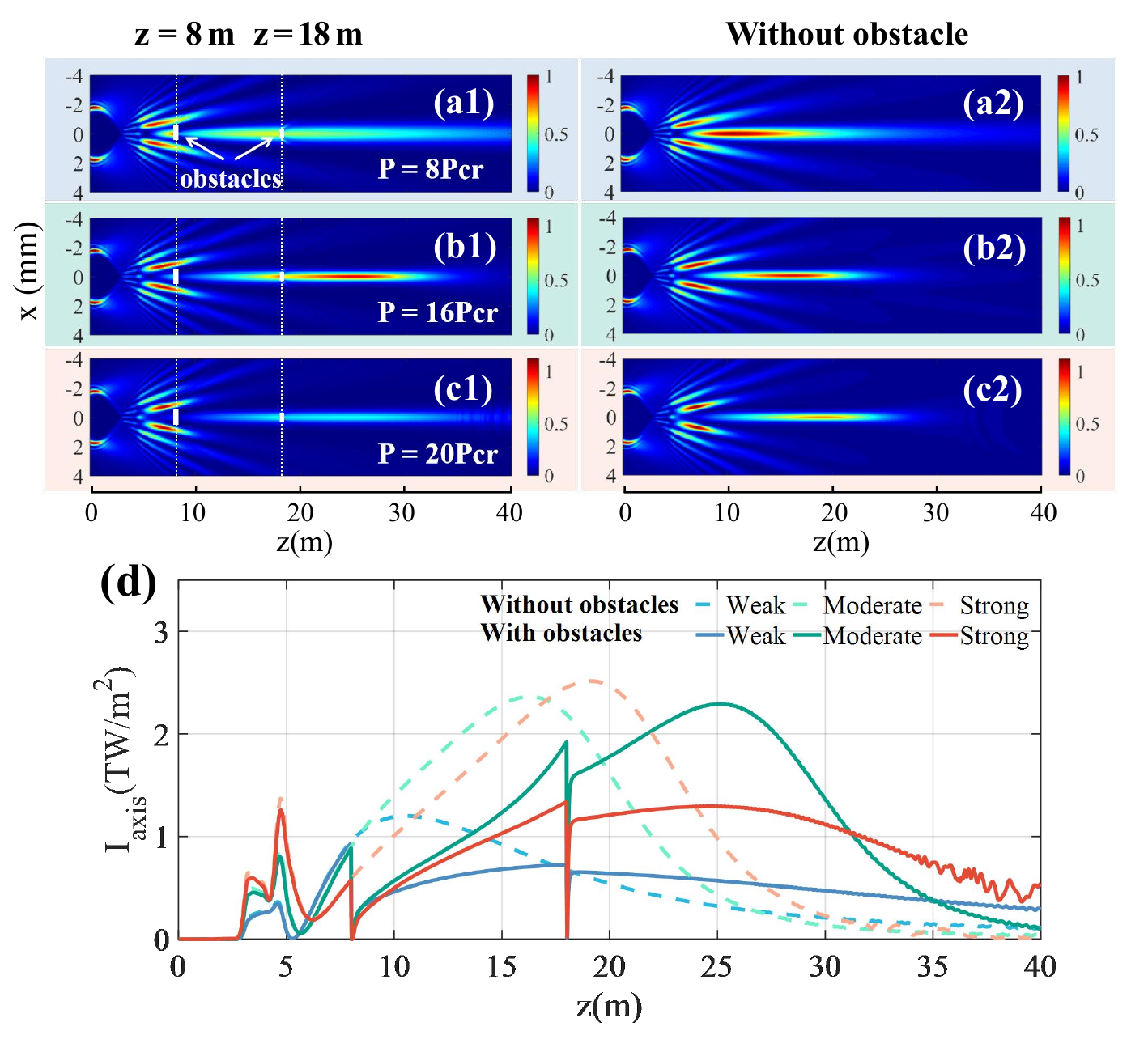}
\caption{Self-healing properties of an EAB with various Kerr nonlinearity when two obstacles are placed on the beam axis. Longitudinal intensity distribution in the $x-z$ plane when (a) $P_{\textup{in}} = 8P_{\textup{cr}}$, (b) $P_{\textup{in}} = 16P_{\textup{cr}}$, and (c) $P_{\textup{in}} = 20P_{\textup{cr}}$. (a1-c1) are obstacles at propagation distance of $8\textup{m}$ and $22\textup{m}$, respectively, and (a2-c2) have no obstacles. (d) Comparison of the on axis intensity at $P_{\textup{in}} = 8P_{\textup{cr}}$ (blue line), $16P_{\textup{cr}}$ (green line), and $20P_{\textup{cr}}$ (red line).}
\label{fig4}
\end{figure}

In this section, we investigate the effects of Kerr nonlinearity on the self-healing properties of EABs with two obstacles along the beam axis. Thus, two obstacles with diameters of $\SI{0.60}{mm}$ and $\SI{0.25}{mm}$ were placed at propagation distances of $\SI{8}{m}$ and $\SI{18}{m}$, respectively. A comparison of the longitudinal intensity distribution and on-axis intensity of various Kerr nonlinearities is plotted in Fig.\,\ref{fig4}, and weak, moderate, and strong nonlinearities are represented by different input powers of $8P_{\textup{cr}}$, $16P_{\textup{cr}}$, and $20P_{\textup{cr}}$, respectively.

From Fig.\,\ref{fig4}, it can be observed that EABs exhibits different self-healing abilities with varying input power, resulting from the effect of Kerr nonlinearity. In the case of weak nonlinearity ($P_{\textup{in}} = 8P_{\textup{cr}}$), the EAB propagates a longer distance after blocking compared with the unobstructed case, but with a lower on-axis intensity (Fig.\,\ref{fig4}(a) and blue line in Fig.\,\ref{fig4}(d)). With moderate nonlinearity ($P_{\textup{in}} = 16P_{\textup{cr}}$), a greater propagation distance is achieved with a minor change in the on-axis intensity after being disturbed (Fig.\,\ref{fig4}(b) and the green line in Fig.\,\ref{fig4}(d)). This result can be explained by the fact that a higher on-axis beam intensity induces a larger refractive index change due to the Kerr effect than under weak nonlinearity, resulting in a greater energy flow to the beam axis. As a consequence, the on-axis beam of the EAB under moderate nonlinearity has a longer propagation distance and maintains high intensity after being blocked by the two obstacles. However, the emission of high-energy solitons by the EAB under strong nonlinearity ($P_{\textup{in}} = 20P_{\textup{cr}}$) results in a lower on-axis intensity, thereby reducing the self-focusing effect. Thus, the on-axis intensity declines rapidly under strong nonlinearity after being disturbed by two obstacles on the beam axis (Fig.\,\ref{fig4}(c) and the red line in Fig.\,\ref{fig4}(d)). Therefore, moderate nonlinearity demonstrates a stronger on-axis intensity and a greater self-healing ability compared with both weak and strong nonlinearities when two obstacles are situated on the beam axis.

\subsection{Multiple random obstacles}
\begin{figure}[tp]
\centering\includegraphics[width=0.95\columnwidth]{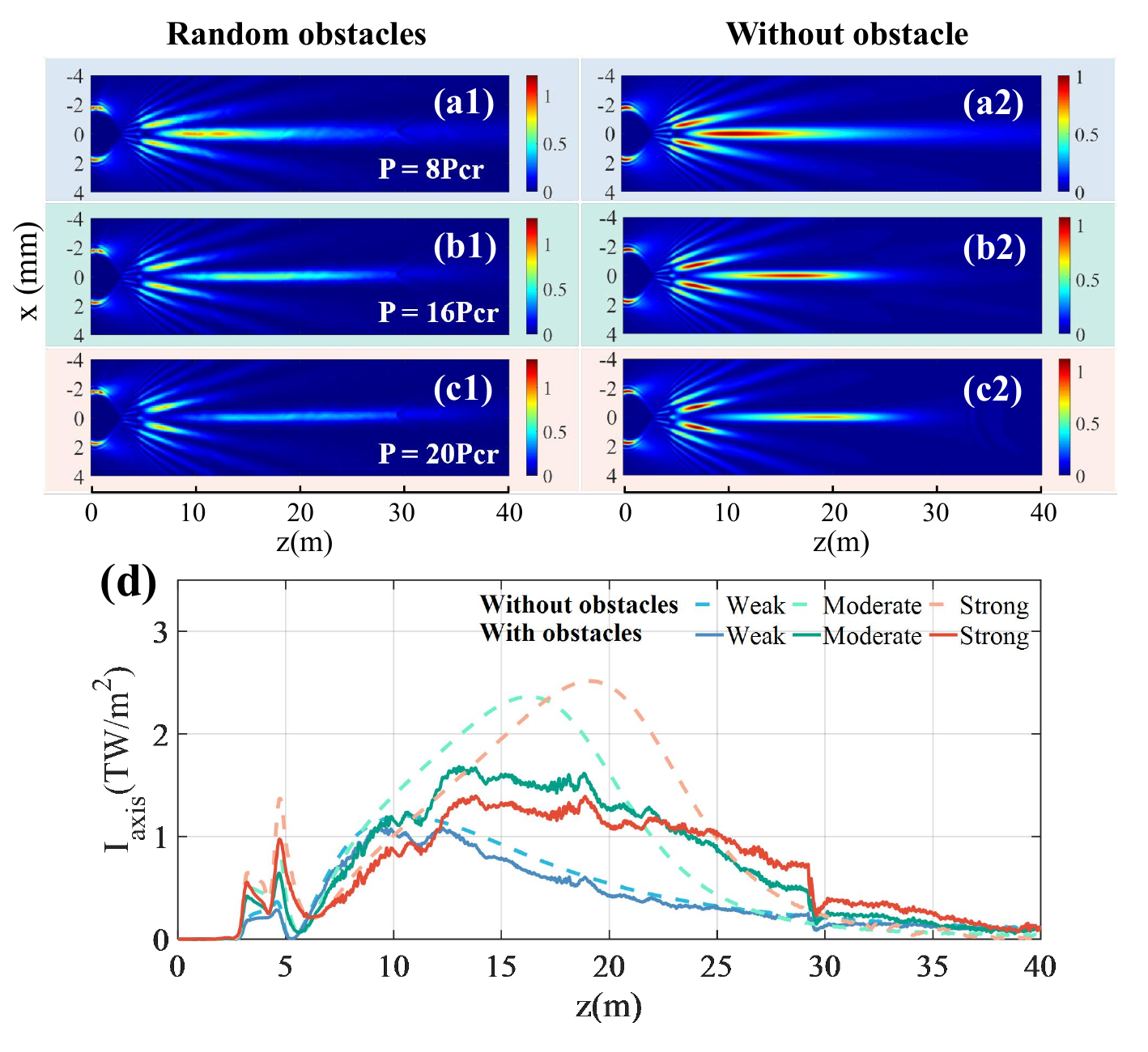}
\caption{Self-healing properties of an EAB with various input powers in multiple random and uniform obstacles media. Longitudinal intensity distribution in the $x-z$ plane when (a) $P_{\textup{in}} = 8P_{\textup{cr}}$, (b) $P_{\textup{in}} = 16P_{\textup{cr}}$, and (c) $P_{\textup{in}} = 20P_{\textup{cr}}$. (a1-c1) are in multiple random and uniform obstacles media, and (a2-c2) have no obstacles. (d) Comparison of the on axis intensity at $P_{\textup{in}} = 8P_{\textup{cr}}$ (blue line), $16P_{\textup{cr}}$ (green line), and $20P_{\textup{cr}}$ (red line).}
\label{fig5}
\end{figure}

To further examine the self-healing properties of EABs in the context of challenging underwater conditions, we introduce numerous small particles distributed uniformly and randomly into the model to simulate the propagation of EABs in turbid water. A total of $4,000$ randomly distributed particles, each with a diameter of \SI{0.3}{mm}, are placed within a propagation range of $\SI{60}{mm} \times \SI{60}{mm} \times \SI{40}{m}$. Additionally, only the absorption of the small particles on the beam energy is considered. The longitudinal intensity distribution and on-axis intensity of the EABs in the presence of multiple obstacles versus the absence of obstacles are illustrated in Fig.\,\ref{fig5}.

As shown in Fig.\,\ref{fig5}, it can be found that the longitudinal intensity distributions are similar to those without obstacle disturbance, except for a small decay in the on-axis intensity. In other words, the longitudinal propagation distributions of the EABs are robust to various small particles. By comparing the on-axis intensity at different input powers, it can be observed that the on-axis intensity decreases less in the case of weak and moderate nonlinearities than strong nonlinearity, indicating a higher self-healing property.

\section{Optimal power of EABs for underwater propagation}
\label{section 5}

\begin{figure*}[tp]
\centering\includegraphics[width=2\columnwidth]{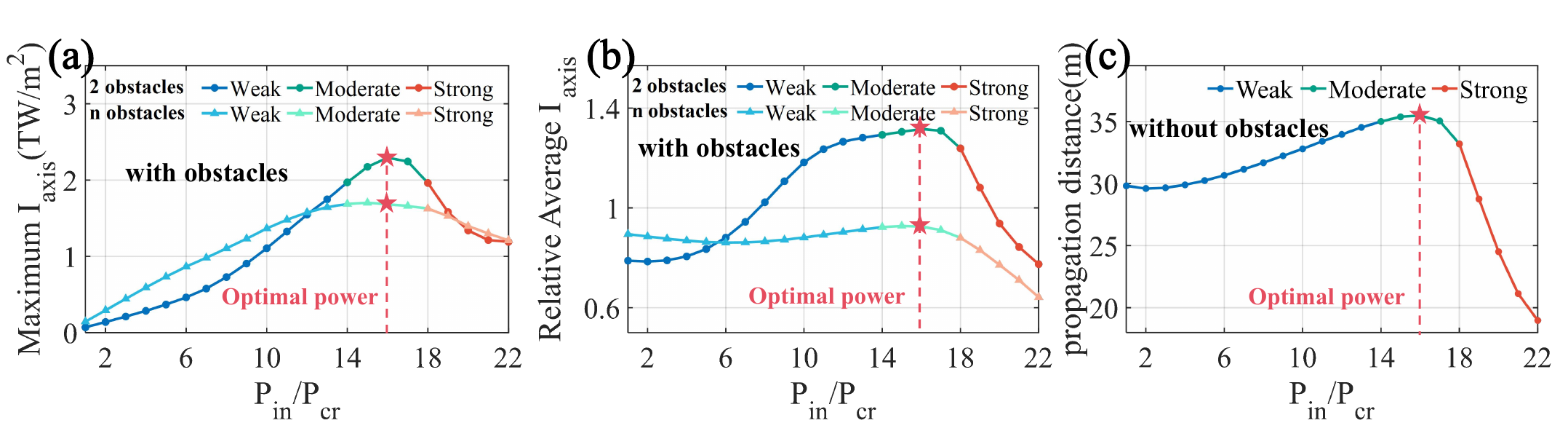}
\caption{Optimal input power. (a) Maximum $I_{\textup{axis}}$ and (b) Relative Average $I_{\textup{axis}}$ versus the relative input power $P_{\textup{in}}/P_{\textup{cr}}$ with 2 obstacles on axis(corresponding to Fig.\,\ref{fig4}) and multiple random obstacles (corresponding to Fig.\,\ref{fig5}). (c) Propagation distance versus the relative input power $P_{\textup{in}}/P_{\textup{cr}}$ without obstacles (corresponding to Fig.\,\ref{fig2}(c)).  }
\label{fig6}
\end{figure*}

The preceding analysis of the nonlinear propagation dynamics of EABs reveals that at moderate nonlinearity, EABs can propagate longer distances in pure water and maintain a higher intensity after being blocked than at weak and strong nonlinearities. To explore whether an optimal input power can achieve the highest beam quality in both pure water and obstacle environments, we quantitatively analyze the on-axis intensity variation of EABs in obstacle environments. To this end, we define the Relative Average $I_{\textup{axis}}$ (RAI) to describe the influence of obstacles on the on-axis intensity of EABs, highlighting their self-healing properties.

\begin{equation}
\textup{RAI} = \frac{\int_{z_1}^{z_2} I_{\scriptscriptstyle \textup{axis}}^{\scriptscriptstyle \textup{obstacle}} (z) dz/(z_2-z_1)}{\int_{z_1}^{z_2} I_{\scriptscriptstyle \textup{axis}}^{\scriptscriptstyle \textup{0}} (z) dz/(z_2-z_1)},
\end{equation}

where $I_{\scriptscriptstyle \textup{axis}}^{\scriptscriptstyle \textup{obstacle}} (z)$ and $I_{\scriptscriptstyle \textup{axis}}^{\scriptscriptstyle \textup{0}} (z)$ represent the on-axis intensity as a function of the propagation distance $z$; and $\int_{z_1}^{z_2} I_{\scriptscriptstyle \textup{axis}}^{\scriptscriptstyle \textup{obstacle}} (z) dz/(z_2-z_1)$ and $\int_{z_1}^{z_2} I_{\scriptscriptstyle \textup{axis}}^{\scriptscriptstyle \textup{0}} (z) dz/(z_2-z_1)$ represent the average intensity of the on-axis beam at propagation distance from the initial plane $z_1 = \SI{0}{m}$ to $z_2 = \SI{40}{m}$, both with and without obstacles, respectively. Specifically, when RAI values are greater than 1, it indicates that in obstacles environments, the average on-axis intensities of the beam increase after encountering obstacles. 

Figure \ref{fig6} presents a comprehensive analysis of the beam quality of EABs with various input powers in different scenarios, both with and without obstacles. In the presence of obstacles (either two or multiple), both the Maximum $I_{\textup{axis}}$ and RAI initially increase with the input power and then decrease, reaching their maximum values at $P_{\textup{in}} = 16P_{\textup{cr}}$, as shown in Fig.\,\ref{fig6}(a) and (b). Thus, by combining the analysis of the on-axis intensity variation of the beam propagation with obstacles and the changes in the propagation distance without obstacles (as shown in Fig.\,\ref{fig6}(c)), it is determined that there is an optimal input power $P_{\textup{opt}} = 16P_{\textup{cr}}$, at which the Maximum $I_{\textup{axis}}$, RAI, and propagation distance all reach their optimal values. This optimal power can be explained as follows: the self-focusing effect on the beam axis compensates for the diffraction-induced beam spreading to the greatest extent possible at the input power of $16P_{\textup{cr}}$, resulting in the optimal propagation distance and beam stability at this point. In other words, the beam achieves the greatest propagation distance in pure water and the highest beam stability in obstacle environments when the input power is the optimal power.

\section{Conclusion}
\label{section 6}
In this study, we investigated the effect of Kerr nonlinearity on the propagation characteristics and self-healing properties of EABs, indicating that EABs can propagate over extended distances in water. By varying the input power we demonstrated that the EABs exhibit the greatest peak intensity and propagation distance under moderate nonlinearity. By studying the impact of Kerr nonlinearity on the self-healing property, we also showed that EABs under weak and moderate nonlinearity are more robust than those under strong nonlinearity in obstacle environments. Finally, we proposed an optimal power that would enable EABs to achieve the greatest propagation distance and exhibit the highest self-healing property in both pure and obstacle environments. Our work provides a comprehensive analysis on the propagation dynamics and self-healing properties of EABs,  which is helpful to understand the behaviors of EABs propagating in nonlinear media. Therefore, applications in underwater wireless optical communication are expected for EABs, where the long-distance propagation and self-healing properties are conducive to underwater propagation.

\section*{Acknowledgments}

This work was supported by the National Natural Science Foundation of China (Grant No. 62027822).

\bibliography{reference}

\end{document}